\documentclass[conference]{IEEEtran}
\usepackage{cite}
\usepackage{amsmath,amssymb,amsfonts}
\usepackage{algorithmic}
\usepackage{graphicx}
\usepackage{textcomp}
\usepackage{xcolor}
\def\BibTeX{{\rm B\kern-.05em{\sc i\kern-.025em b}\kern-.08em
    T\kern-.1667em\lower.7ex\hbox{E}\kern-.125emX}}
\begin{document}
\title{A review on Deep Neural Network for Computer Network Traffic Classification}

\author{\IEEEauthorblockN{Md. Ariful Haque, Dr. Rajesh Palit}
\IEEEauthorblockA{\textit{Dept of  Electrical and Computer Engineering} \\
\textit{North South University}\\
Dhaka, Bangladesh \\
\{ariful.haque1, rajesh.palit\}@northsouth.edu}
}
\maketitle
\maketitle

\begin{abstract}

\textcolor{purple}{Focus on Deep Neural Network based malicious and normal computer Network Traffic classification. (such as attacks, phishing, any other illegal activity and normal traffic identification).}
In this paper, the main idea is to review, existed Neural Network based network traffic classification. Which indicates intrusion activity classification and detection. It is very important to classify network traffic to safeguard any system, connected to computer network. There are a variety of NN architecture for it, with different rate of accuracy. On this paper we will do relative compression among them.
\end{abstract}


\section{Introduction}
Now a days, computer network plays an essential role in our world economy and society. We cannot think any moment without being connected to network. Based on network many Internet of Things (IoT) systems also developed~\cite{castellani2010architecture}. There are modern power systems, for home and others, connected to computer network, to be controlled power flow from power grid company~\cite{boardman2015intelligent}\cite{hartig1996subscriber}. On the other hand, today’s modern vehicular systems also embeds with a number of computer devices called Electronic Control Unit (ECU), which communicates each other over in-vehicle networks and internet to facilitate advanced automotive applications like auto-driving service and many more~\cite{yu2008integrated}. Mainly, with the development of network techniques and science technologies, IT industry has expanded greatly. Almost organizations such government, enterprises, banking system .etc and as well as personal users also getting depended on computer networks more and more. Such services demands high attentions on data security and integrity. Additionally, our network architecture that us use world wide, has by born architectural defect on its different layer~\cite{zhang1987designing}. As our network architecture and protocols was built on early ages and at that point of time no one thought about those security holes, that we are facing now. There are intruders~\cite{staniford1995holding} (inside or outside), hackers, spammers and may more, who are simultaneously trying to break or crack our network system and protocols to gain access with criminal intention.
 
As our network systems has packets~\cite{buckman2006system} on its different layer, all kind of network activity transmits on packet basis. In general, incoming and outgoing activities of those data packets are known as network traffic. To monitor or detect any unusual activity on computer network, "classifying network traffic" approach is widely used. Simply, by observing if we can distinguish normal network traffic from all other, then all network attacks can be stopped beforehand. And this is the key idea of any Intrusion detection and prevention System. It classifies and distinguishes normal traffic from all others. Traditional network traffic classification is done in misused based detection approach~\cite{roesch1999snort}~\cite{paxson1999bro}. Where network attacks  are identified by using a predefined attack signatures. There are some generic algorithms for it like\cite{paliwal2012denial}\cite{dave2014improved}\cite{akbar2011implementing}. Though this traditional approach is quite good and detects known attacks effectively. Moreover this approach has remarkable disadvantages. It is effective only for known attacks and does nothing for unknown attacks. As attackers building new attack tools and ways, most likely on daily basis, it requires simultaneous update on its attack signature data-set, to be up to date. Which indicates that, new attack cannot be handled before deploying and this is really a dangerous and expensive security hole. Some attacks like zero-day exploit, worm etc. uses polymorphism that delays generation of attack signatures~\cite{bilge2012before}~\cite{aziz2011computer}~\cite{fogla2006polymorphic}. All of this factors put a question mark on this traditional traffic classification. To fill that gap, a new anomaly based traffic classification comes on the stream, which includes Deep Neural Network. According to some Cyber security expert, new attacks are just a variance of some known attack and by the similarity of their parameters. It is possible to detect these variations~\cite{araujo2010identifying}. Deep Neural Network does excellent job on this type of scenario. Another good thing about this approach is, it yields more than 90\% of accuracy and also covers new type of attacks.

\section{Deep Neural Network}
A deep Neural Network consist of a processing element named neuron. At first it produced in 1943 by the neurophysiologist Warren McCulloch and a logician Walter, Pitts~\cite{mcculloch1943logical}. It aligns in group, takes inputs and produces preferable outputs as per model design~\cite{liu2017survey}. It has three phase; training, validation and testing. On a typical Neural network, every neuron has its own bias and wight. Based on them, every neuron works on group basis and produces a predicted result. Here at first, model gets trained with help of training data-set and it makes necessary changes on all of it's parameters. For network traffic classification, model was trained using previous network traffic history (real routing packets). After validation and testing it gets deployed on firewall, IDS, IPS or other respective applicable systems. With the power of DNN models, those system can distinguish any normal or unusual traffic, as well as unknown traffic pattern. Because of their generalization feature, neural networks are able to work with imprecise and incomplete traffic pattern. Another great advantage is analyzing speed, which is faster then old classification techniques.~\cite{ahmad2008performance}

\section{Used Traffic types}
There are numerous type of network traffic. For DNN training, more or less every successful Technique used this 5 type of traffic data during their model building.:Normal, DOS, U2R, R2L and Probing
\begin{itemize}
\item 	\textbf{Normal Traffic}
This is as usual expected traffic type. Normal TCP, UDP, ICMP or any other certified protocol with valid purpose. 

\item 	\textbf{DOS}
Denial of Service (DoS) attack. It occurs when attacker attempts to overload, disable or disrupt any system resource by sending tremendous amount of traffic on a short duration of time\cite{park2001effectiveness}. In general they try to occupy all available bandwidth. As a result legitimate users are prevented access to network resources.

\item 	\textbf{R2L}
Remote-to-Local attack. Here attacker tries to gain access to a system as local user without being a valid user of that system\cite{sabhnani2003kdd}. They send illegal packet for it. Example: Multihop, Ftp-write.

\item 	\textbf{Probe Attack}
Here attacker tries to get information about a target computer network. Main aim is to find some vulnerability on that network. They track the network topology of target system and tries to discover running services on that network\cite{kondo1978multiple}. Some attack tools Example: Nmap, Ipseewp, Satan.

\item 	\textbf{U2R}
In User-to-Root attack. Using some common techniques like Social engineering, password sniffing, spoofing, SQL injection, cross script injection etc. hacker tries to get root access of target system\cite{beghdad2009efficient}. Attack tools Example: Rootkit, Perl.

\end{itemize}

\section{DNN Techniques}
\subsection{Technique-1}
One of the approach for network traffic classification by DNN was done by Basant et al. in 2016~\cite{subba2016neural}. They train the model using NSL-KDD~\cite{tavallaee2009detailed} data-set. Which consist of 41 different features. On their architecture, they have one input layer, one output layer and only one hidden layer. It was feed forward network and Sigmoid was used there. Their model can distinguish 5 type of traffic which are: Normal, DOS, U2R, R2L and Probing. Among them, their model can detect DOS attack up to maximum 99.23\% with 97.39\% accuracy. And U2R attack minimum 72.24\% with 97.67\% accuracy. Their Normal traffic detection rate is 88.48\% with 94.74\% accuracy.

\subsection{Technique-2}
Another model was proposed by MJ kang et al. in 2016~\cite{kang2016novel} and it was for automobile. They used Controller Area Network (CAN) packet~\cite{pazul1999controller} for detection which is 64 bit long. Their idea was to detect only normal traffic. At first they trained their model using known attack and after deploying, anything unknown will be considered as new attack and will be recorded. Their proposed architecture consist of 64 dimensional input layer, multiple hidden layer of fixed 64 dimension and 1 dimensional output layer. They used ReLU here. For hidden layer they used three different arrangement which yields three different accuracy percentage. For 3 hidden layer arrangement, attack detection rate is 97.6\% with 95\% accuracy. 5 and 7 hidden layer yields respectively, attack detection rate is 99.8\% with 95.7\% accuracy and attack detection rate is 99.8\% with 99.9\% accuracy. Which indicates, 7 hidden layer arrangement is best among them.

\subsection{Technique-3}
RS Naoum et al. in 2012~\cite{naoum2012enhanced} presented another Network traffic classification model. Their model consist of one hidden layer with 26 neurons. Detects 5 different traffic, therefore 5 neuron on output layer. The classes are Normal, DOS, U2R, R2L and Probing. Instead of traditional Backpropagation, the used RPROP~\cite{riedmiller1993direct}(resilient propagation) here. Which depends on the sign of derivative, not its value. It uses only partial derivative of the error. Therefore it is faster than normal Backpropagation algorithms. They train their model with different number of neuron on hidden layer. They started training with 10 neurons and ends with 32. Among them 26 neuron yields best result with 94.7\% total delectation rate. Additionally, the model can detect maximum 99.8\% of probing, U2R attack minimum 54.1\%. Their normal traffic detection rate is 84.3\%.

\subsection{Technique-4}
Another remarkable work on this field is done by M Pradhan, et al. in 2012~\cite{pradhan2012anomaly}. Which yields maximum of 88\% detection rate with 100\% accuracy. The accuracy level is really a good number. Their model can detect 5 type of traffic called Normal, DOS, U2R, R2L and Probing. they used DARPA Intrusion Detection Evaluation dataset~\cite{debar1992neural} from 1998. There are 7 different weeks traffic log on that dataset. For their model they used only 3rd weeks traffic log. They tried with 2 to 6 different hidden layer. Among them, overall 6 hidden layer yields better.

\subsection{Technique-5}
Another approach is done by TA Tang et al. in 2018~\cite{tang2018deep}. They used Recurrent Neural Network. NSL-KDD~\cite{tavallaee2009detailed} as their dataset and it is a balanced enough. They used a Nadam optimizer~\cite{dozat2016incorporating} and a mean squared error (MSE) for their model. In addition, they added L 1-regularization to prevent over-fitting during the training phase. And they ends up with maximum 89\% accuracy.

\subsection{Technique-6}
One Convolutional Neural Network Based approach was proposed by W Wang et al. in 2017~\cite{wang2017malware}. They used LeNEt-5~\cite{lecun1995learning} architecture. For dataset, they created their own dataset USTC-TFC2016 from~\cite{ctudataset} and IXIA BPS~\cite{ixiadataset} , which is different then typical dataset, as it has raw traffic data. They converted data into 28 X 28 X 1 dimensional gray image and normalized to [0,1] from [0,255]. They used max polling in between convolution layer and Softmax in the end of Fully connected layer according to LetNET-5~\cite{lecun1995learning}. They designed three type of classifier fro 20 different class. One is Binary classifier. It classifies into Normal traffic and Malware traffic. Again malware and normal classifies 10 independent classes each. The second classifier is 20-class classifier. It directly classifies into 20 classes, including normal and malware type. Finally they gained average 99.41\% accuracy. Individually Binary classifier yields 100\% accuracy and 20-class Classifier yields 99.17\% accuracy. 

\subsection{Technique-7}
LP Dias et al. in 2017~\cite{dias2017using} presented another Network traffic classification model. They used KDDCUP’99~\cite{dias2017using} as their dataset. Which covers Normal, DOS, U2R, R2L and Probing, these 5 type of traffic. There are only one hidden layer. There input layer takes 41 dimensional input. 5 class to detect, therefore 5 node on output layer. On hidden layer they tried 5 (with 20, 30, 40, 50 and 55) different number of neuron. Among them 50 neuron hidden layer edition gives maximum result. For Normal, DOS, Probe, R2L and U2R, accuracy respectively, 99.8\%, 100\%, 100\%, 96.1\% and 51.9\%, with an average classification rate of 99,9\%.

\subsection{Technique-8}
One of the early age model of DNN based network traffic classification was done by J Ryan et al. in 1998~\cite{ryan1998intrusion}. They used standard 3 layer backpropagation architecture. Which 100 dimensional input. 1 hidden layer consist of 30 neurons and classifies 10 claas, thus 10 output neurons. They used 12 days user log for 10 user from University of Texas, as their dataset. Those users belong to a research group of that University. Their model yields 96\% accuracy on detection and 93\% accuracy on classification.

\subsection{Technique-9}
Another technique introduced by C Jirapummin et al. in 2002~\cite{jirapummin2002hybrid}. They used KDD cup 1999~\cite{stolfo1999kdd} dataset. They used Self-organizing map (SOM) on top of DNN. Their DNN model consist of 2 hidden layer with 70 and 12 neuron accordingly. They classified 4 classed therefor 4 neuron on output layer. For back propagation they used the used RPROP~\cite{riedmiller1993direct}(resilient propagation). It reduces their training time significantly. They also used log-sigmoid and tan-sigmoid. They classified 3 type of attack traffic and one normal type. Classes are Normal, Neptune, portsweep and Satan. Their model yield maximum 99.7181\% detection rate of Neptun and minimum 90.2811\% detection of Satan. False alarm rate for Neptun is 0.0591, pretty low. Which indicates better accuracy.

\subsection{Technique-10}
Another standard approach was proposed by I Ahmad et al. in 2008 ~\cite{ahmad2008performance}. They used KDD cup 1999~\cite{stolfo1999kdd} as their dataset. They experimented with 1 to 4 different hidden layer. Among them 2 hidden layer alignment yields better. First hidden layer consist of 41 and second one consist of 9 neuron. The input layer consists of 41 neurons as their dataset contains 41 fields for a network packet. They used sigmoid as their transfer function. They experimented backpropagation with RPROP~\cite{riedmiller1993direct}(resilient propagation). RPROP produced better result.

\section{Discussion and Conclusion}
Those are the mostly popular techniques or methods for classifying Network traffic with help of Neural Network. The data they obtained for training and testing, either from real traffic like~\cite{stolfo1999kdd} or simulated traffic like~\cite{wang2017malware}. Different researchers used different Network architecture as their design, but most of them used single hidden layer and obtained good result. Some used normal NN with 1 to 7 hidden layer, some additionally used RPROP\cite{naoum2012enhanced}\cite{jirapummin2002hybrid}\cite{ahmad2008performance}, some used Convolutional Neural Network~\cite{wang2017malware}, some used Recurrent Neural Network like~\cite{tang2018deep}. They used MATLAB, PlaNet, OPNET NeuralWorks simulators and other personally developed ways. Overall there accuracy level yields up to 100\%~\cite{pradhan2012anomaly}, on their testing set. Though it generates some training overhead, for every case Neural Network produces remarkable result. In real scenario, those model can reduce classification, more specifically detection time. Despite all, Application of Neural Network in computer network classification is an ongoing area and is limited to academic research till now\cite{ahmad2009artificial}. This field worth more works for more accuracy and precision.  
\bibliographystyle{unsrt}
\bibliography{bibliography.bib}

\end{document}